\documentclass[12pt]{article}

\usepackage{scicite}
\usepackage{amsmath}
\usepackage{times}
\usepackage{graphicx}
\usepackage{xcolor}
\usepackage{parskip}
\usepackage{enumitem}%
\usepackage{comment}
\usepackage{caption}
\usepackage{float}
\usepackage{booktabs}
\usepackage{color,soul}
\usepackage{tabularx}
\usepackage{amsmath,amssymb}
\usepackage{ragged2e}
\usepackage{bm}
\usepackage{xr}
\usepackage{filecontents}

\makeatletter
\newcommand*{\addFileDependency}[1]{
  \typeout{(#1)}
  \@addtofilelist{#1}
  \IfFileExists{#1}{}{\typeout{No file #1.}}
}
\makeatother

\newcommand*{\myexternaldocument}[1]{%
    \externaldocument{#1}%
    \addFileDependency{#1.tex}%
    \addFileDependency{#1.aux}%
}

\myexternaldocument{SI}

\newenvironment{sciabstract}{%
\begin{quote} \bf}
{\end{quote}}

\title{Molecular-scale substrate anisotropy and crowding drive long-range nematic order of cell monolayers}

\author
{Yimin Luo$^{1,2}$, Mengyang Gu$^{3}$, Minwook Park$^{4}$, Xinyi Fang$^{3}$,\\ Younghoon Kwon$^{2}$, 
Juan Manuel Urue\~{n}a$^{5}$, Javier Read de Alaniz$^{4}$, \\Matthew E. Helgeson$^{1}$, M. Cristina Marchetti$^{6}$ and Megan T. Valentine$^{2,\ast}$\\
\\
\normalsize{$^{1}$Department of Chemical Engineering}\\
\normalsize{$^{2}$Department of Mechanical Engineering}\\
\normalsize{$^{3}$Department of Statistics and Applied Probability}\\
\normalsize{$^{4}$Department of Chemistry and Biochemistry}\\
\normalsize{$^{5}$BioPACIFIC MIP, California NanoSystems Institute}\\
\normalsize{$^{6}$Department of Physics, University of California, Santa Barbara, USA}\\
\\
\normalsize{$^\ast$To whom correspondence should be addressed; E-mail: valentine@engineering.ucsb.edu}
}

\date{}

\begin{document} 


\baselineskip24pt


\maketitle


\begin{sciabstract} 

The ability of cells to reorganize in response to external stimuli is important in areas ranging from morphogenesis to tissue engineering. Elongated cells can co-align due to steric effects, forming states with local order. We show that molecular-scale substrate anisotropy can direct cell organization, resulting in the emergence of nematic order on tissue scales. To quantitatively examine the disorder-order transition, we developed a high-throughput imaging platform to analyze velocity and orientational correlations for several thousand cells over days. The establishment of global, seemingly long-ranged, order is facilitated by enhanced cell division along the substrate’s nematic axis, and associated extensile stresses that restructure the cells' actomyosin networks. Our work, which connects to a class of systems known as active dry nematics, provides new understanding of the dynamics of cellular remodeling and organization in weakly interacting cell collectives. This enables data-driven discovery of cell-cell interactions and points to strategies for tissue engineering.
\end{sciabstract}



\section*{Introduction}

Active matter comprises systems of agents or particles that individually consume energy from the environment to generate motion and forces, and collectively organize in emergent structures on scales much larger than the individual \cite{ramaswamy2017active,marchetti2013hydrodynamics}. In particular, active nematics \cite{doostmohammadi2018active,giomi2014defect} are collections of elongated, apolar active particles that organize in states of orientational order. Nematic order has been observed ubiquitously in active and living systems, from reconstituted suspensions of cytoskeletal filaments and associated motor proteins \cite{Sanchez2012,decamp2015orientational,Kumar2018} to cell monolayers~\cite{kemkemer2000elastic,duclos2014perfect,saw2017topological,kawaguchi2017topological}, bacterial colonies \cite{thutupalli2015directional}, and even on the scale of entire organisms \cite{maroudas2021topological}. 

The nematic arrangement of cells in biological systems appears to serve key biological functions, such as driving the expansion of bacterial colonies~\cite{basaran2022large}, controlling cell extrusions and multilayer formation in confluent tissue~\cite{saw2017topological,kawaguchi2017topological,copenhagen2021topological}, and providing an underlying organizational structure for morphogenetic processes~\cite{maroudas2021topological}. This realization has motivated efforts to develop \textit{in vitro} techniques to control cell organization, which are important both as platforms for controlled fundamental studies as well as for tissue engineering. Established methods include patterning of the topography \cite{turiv2020topology}, stiffness \cite{islam2016collagen} and mechanical stretching \cite{chen2018role} of the substrate. These methods control the orientation at the level of individual cells such that even isolated cells can be sufficiently polarized to follow a preferred direction. Thus, they do not allow the study of the role of steric effects or other aligning mechanisms in tuning the onset of nematic order in the cell collective.

As a complement to topographically modified substrates, recent work~\cite{martella2019liquid} showed that myoblasts cultured on uniform, flat substrates made of liquid crystal elastomers (LCEs) developed nematic order, but only at the collective level. Although this result hinted at the presence of a density-driven isotropic-nematic transition in the orientational order of cells, the mechanisms through which both the structure of the substrate and cell proliferation drive alignment remain largely unclear. This is in part because of a lack of quantitative, time-resolved analysis of dynamical trajectories for a statistically meaningful number of cells that would inform such mechanistic insights. 

In this work, we develop methods for the high-resolution tracking and quantitative analysis of several thousand shape anisotropic cells over days, and demonstrate experimentally that elongated apolar active units can order through steric effects. Specifically, we study the organization and motility of human dermal fibroblasts (hdFs), chosen because of their shape anisotropy. By growing these weakly interacting cells on topographically flat LCE substrates, we can examine the interplay between cell crowding and molecular-level guidance from the substrate in controlling the establishment of orientational order. Earlier work finds that when the substrate is isotropic, steric effects alone establish domains of local orientational order on scales of 10-15 cells~\cite{duclos2017topological,garcia2015physics,kawaguchi2017topological,li2017mechanism,duclos2014perfect}. These domains are randomly oriented and the orientational order does not persist on the scale of the entire tissue. By contrast, we find that a nematic substrate provides a direction of broken symmetry and drives the establishment of nematic order over large ($\sim$ millimeter) scales. In this study, we develop techniques to visualize and analyze cellular dynamics at a high space and time resolution for thousands of cells by recording long trajectories over a very large field of view. Using this method, we show that the establishment of global nematic order occurs in a three-step process. With increasing cell density, the system transitions from (i) a disordered state where individual cell trajectories are unaffected by substrate alignment to (ii) an intermediate state where chaotic bands of aligned cells coexist with disordered regions, while the system remains isotropic at the global scale, and finally (iii) an ordered nematic state where the tissue exhibits long-range order on millimeter scales.

Nematic order has been studied extensively in confluent monolayers of epithelial cells in previous works \cite{garcia2015physics,saw2017topological}, where the anisotropy of individual cells is very small and nematic order is believed to arise not from steric effects, but from the anisotropy of forces transmitted through strong cell-cell interactions \cite{trepat2009physical,brugues2014forces,balasubramaniam2021investigating}. In our system, by contrast, cells are highly anisotropic even when isolated and weakly interacting, and alignment is driven by crowding and steric repulsion. On nematic substrates, we identify a strong correlation between the axis of cell division and the direction of substrate alignment, which suggests that directed cell division may play a role in the establishment of order on the tissue scale. 

Our system can be considered an experimental realization of a dry active nematic~\cite{peruani2006nonequilibrium,baskaran2008enhanced}, which has been studied extensively via large scale simulations~\cite{chate2020dry}. Here ``dry'' refers to the situation where the dominant dissipation mechanism is frictional coupling to a substrate, while viscous dissipation from cell-cell interaction and hydrodynamic couplings mediated by the surrounding medium are negligible~\cite{marchetti2013hydrodynamics}. An important distinction, however, is that while in the numerical models nematic order appears spontaneously, here  order is externally biased by the direction of substrate alignment, which is  essential for the establishment of global order. Finally, the large experimental datasets and statistical analyses we develop not only shed light on the mechanisms driving cellular reorganization, but provide dynamical information at the single cell level that is required for the calibration and refinement of physical and machine learning models by data inversion in future works.

\section*{Results}

\subsection*{Nematic order within the LCE directs cell alignment} 

\begin{figure}[th]
    \centering
    \includegraphics[width=0.9\textwidth]{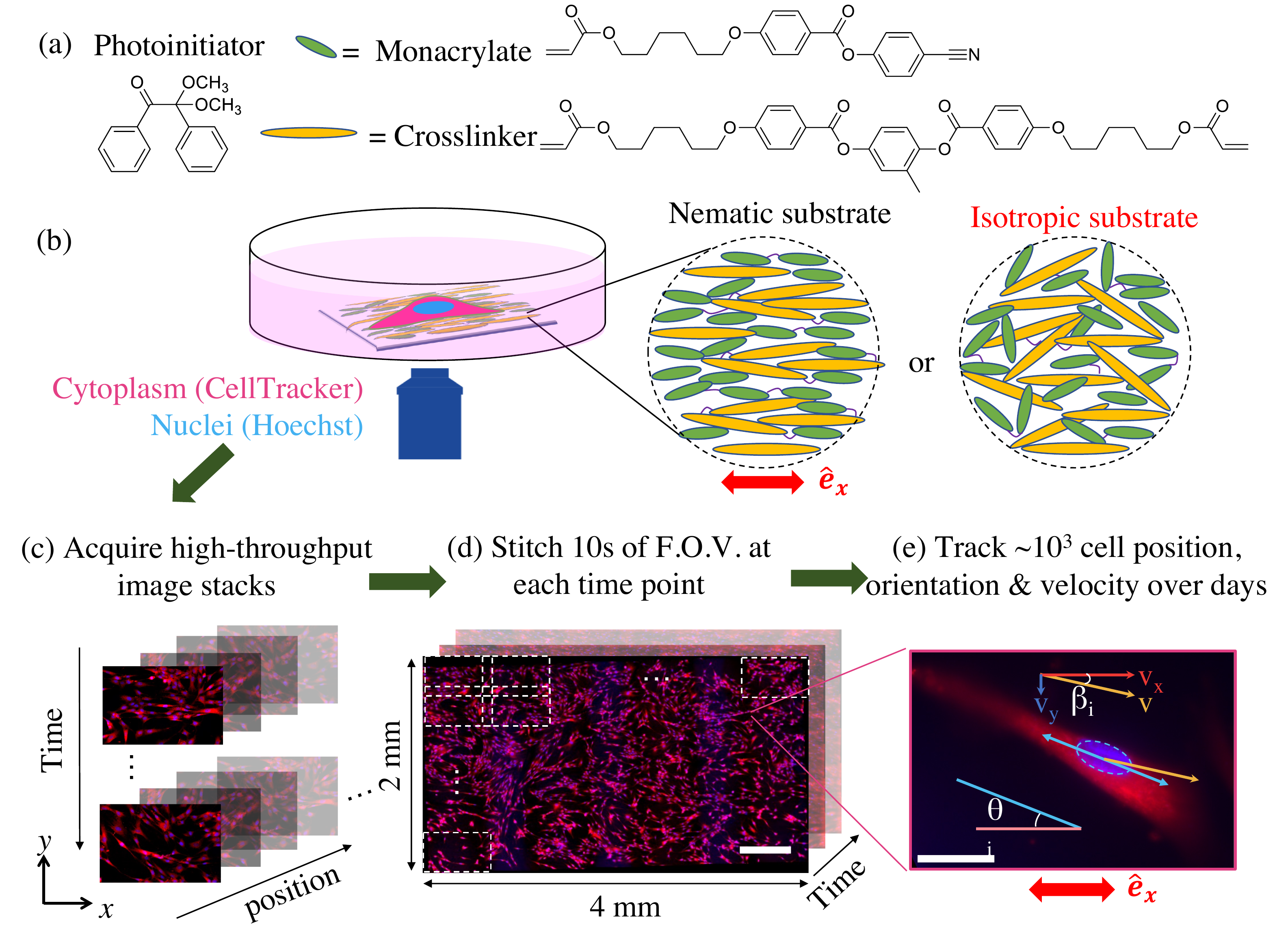}
    \caption{Experimental approach. (a) Chemical structures of the crosslinker (RM82), monoacrylate (RM23) and photoinitiator. The full chemical details and substrate characterization data are provided in the Material and Methods Section. (b) Schematics of cell (not to scale) placed on a substrate with nematic or isotropic molecular structure. Cells are labeled in both the cytoplasm (CellTracker, red) and nucleus (Hoechst, blue) channels. (c-d) A high-throughput automated microscope stage acquires 10s of images at every time point, which are stitched together (dotted rectangles) to allow reconstruction of 10$^3$ cell trajectories over the time span of days. The scale bar is 500 $\mu$m. The dotted rectangle represents one field-of-view captured by a 20x objective. (e) A close-up image of a single cell, with nucleus orientation labeled in blue. The velocity vector (with $x$- and $y$-components) is shown in yellow (red and blue), and the angle $\beta$ denotes the velocity orientation with respect to the substrate. The scale bar is 100 $\mu$m. The substrate alignment direction is parallel to $\hat{\mathbf e}_x$ unless otherwise specified. }
    \label{fig:substrate}
\end{figure}

\begin{figure}
    \centering
    \includegraphics[width=0.6\textwidth]{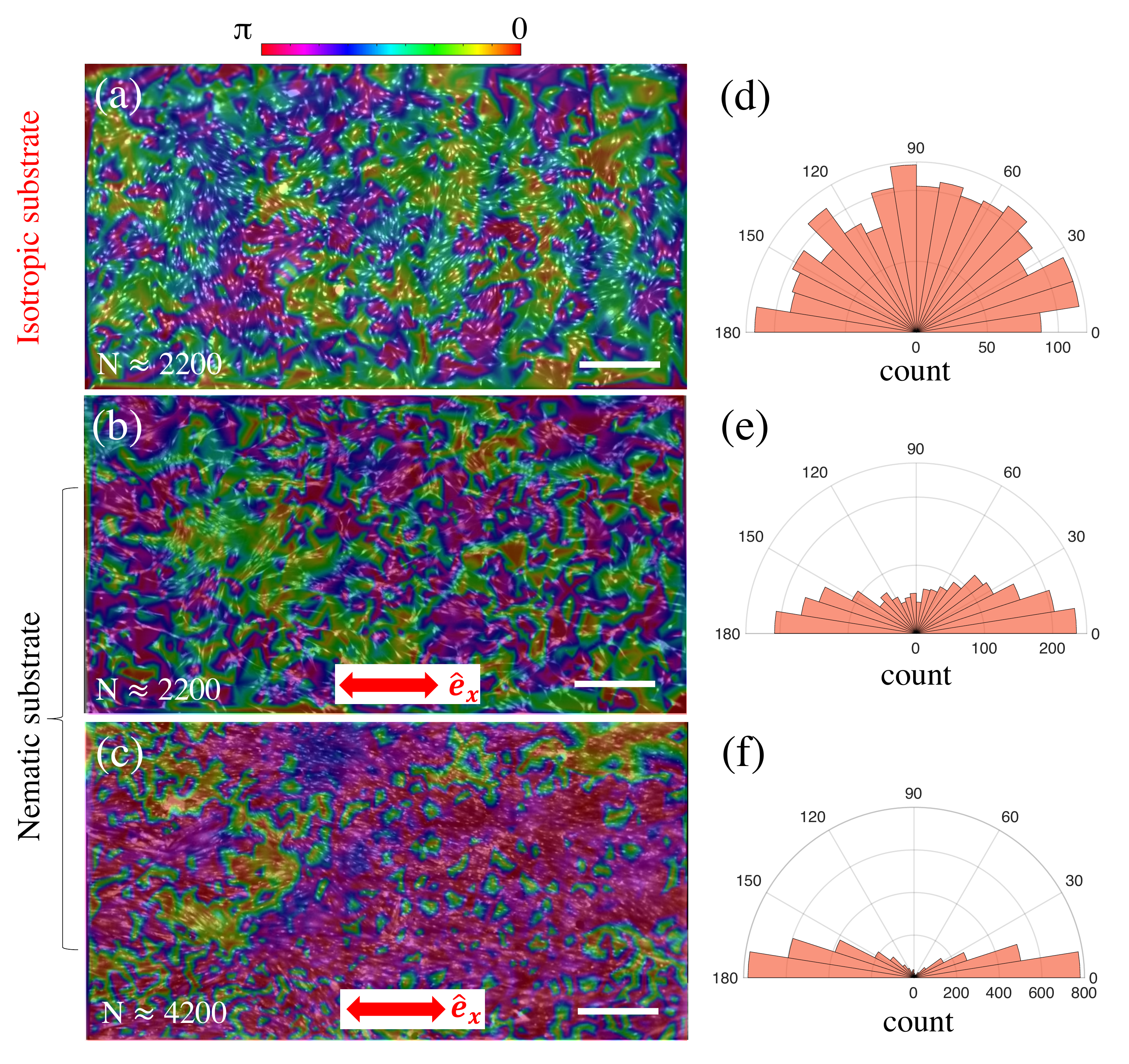}
    \caption{Snapshots of cell configurations on (a) an isotropic substrate with $\approx$ 2200 cells and nematic substrates with (b) $\approx$ 2200 cells and (c) $\approx$ 4200 cells.  The images are obtained by merging the nucleus and cytoplasm channels.  The color scale  denotes the nucleus  orientation.  
    (d-f) show the corresponding polar histograms of the nucleus orientations, with respect to an arbitrary direction (d) and to the direction  $\hat{\mathbf e}_x$ of nematic order of the substrate (e,f). The scale bars are 500 $\mu$m.}
    \label{fig:aligned_vs_isotropic}
\end{figure}

Leveraging our ability to visualize and analyze the dynamics of thousands of cells at a high space and time resolution (Fig. \ref{fig:substrate}), we observed a markedly different organization of cells on isotropic and nematic LCE substrates, indicating that hdF cells are sensitive to the molecular alignment of the polymer  film at the nanoscale, and that they use this molecular information to control their orientation within the cell monolayer. Snapshots of cell orientation  on isotropic (Fig.~\ref{fig:aligned_vs_isotropic}a) and nematic (Fig.~\ref{fig:aligned_vs_isotropic}b,c) substrates illustrate the role of both substrate alignment and cell density in controlling  cell ordering. Cell nuclei are elongated and their orientation is correlated with cell elongation (Fig. \ref{fig:nucleus_body_alignment}). The cell orientation is then defined by the angle $\theta_i$ between the long axis of the nucleus of cell $i$ and a fixed direction $\mathbf{\hat e_x}$. For nematic substrates, $\mathbf{\hat e_x}$ coincides with the direction of LCE alignment, whereas for  isotropic substrates $\mathbf{\hat e_x}$ represents an arbitrary direction. The  angles $\theta_i$'s are color-coded in the images. The corresponding angular distributions are shown as polar histograms in Fig. \ref{fig:aligned_vs_isotropic}d-f.
At a high enough density, cells cultured on an isotropic substrate form locally aligned domains (Fig. \ref{fig:aligned_vs_isotropic}a), 
with a nearly uniform distribution of $\theta_i$ across the entire monolayer (Fig. \ref{fig:aligned_vs_isotropic}d). In contrast, cells cultured on nematic substrates at similar density preferentially align with the substrate nematic orientation $\hat{\mathbf e}_x$ (Fig. \ref{fig:aligned_vs_isotropic}b), with a strongly anisotropic  distribution of $\theta_i$ peaked at 0 or $\pi$ (Fig. \ref{fig:aligned_vs_isotropic}e). At higher cell density on isotropic substrates, the angle distributions become more asymmetric and the domain size increases (Fig. \ref{fig:aligned_vs_isotropic}c). 

To quantify the spontaneous nematic order of the cells and nematic order induced by the substrate, we introduce two order parameters: the cell-substrate order parameter, $S_{cs}$, generalized from the Landau-de Gennes theory of liquid crystals \cite{babakhanova2020cell,turiv2020topology, martella2019liquid}, and the cell-cell order parameter, $S_{cc}$. These are given by
\begin{align}
    &S_{cs}= \langle 2\cos^2\theta_i -1 \rangle\;,
     \label{eq:2d_order_param}\\
    &S_{cc}= \langle
   2\cos^2\theta_{ij} -1 \rangle\;,
    \label{eq:cc_order_param}
\end{align}
where 
$\theta_{ij}=\theta_i-\theta_j$ is the angle between the orientation of cell $i$ and cell $j$. The brackets denote an average  over the entire system. A value of $S_{cs} = 0$ indicates no preferential alignment of the cells with respect to $\hat{\mathbf e}_x$, whereas a value of $S_{cs} = 1$ indicates perfect alignment. Similarly, a substantially larger than zero value of $S_{cc}$ denotes nematic alignment due to cell-cell interactions along a direction of spontaneously broken symmetry. It is known that the system size can significantly impact the value of both $S_{cs}$ and $S_{cc}$ (Fig. \ref{fig:box_size}). This highlights the need for the large-scale imaging used in the present experiments. When using the full composite image size of 2 mm $\times$ 4 mm, we found $S_{cs} = 0$, $0.4$ and $0.68$ for the images shown Fig. \ref{fig:aligned_vs_isotropic}a, b, and c, respectively, in agreement with our qualitative observations of orientational order.

\subsection*{The isotropic-nematic transition of the cell monolayer is density-dependent}

Figure~\ref{fig:order_vs_density}a shows the evolution of the cell-substrate order parameter $S_{cs}$ with cell density, with each density value represented by a different color. To probe a wide range of densities, we imaged cells on different days after seeding (seeding densities $\rho_s \approx$ 50 mm$^{-2}$).
Within a single experiment, cell proliferation can also lead to an increase in cell density over the acquisition time ($\sim$ 40 hours). Three key stages are described below:

\begin{figure}
    \centering
    \includegraphics[width=0.8\textwidth]{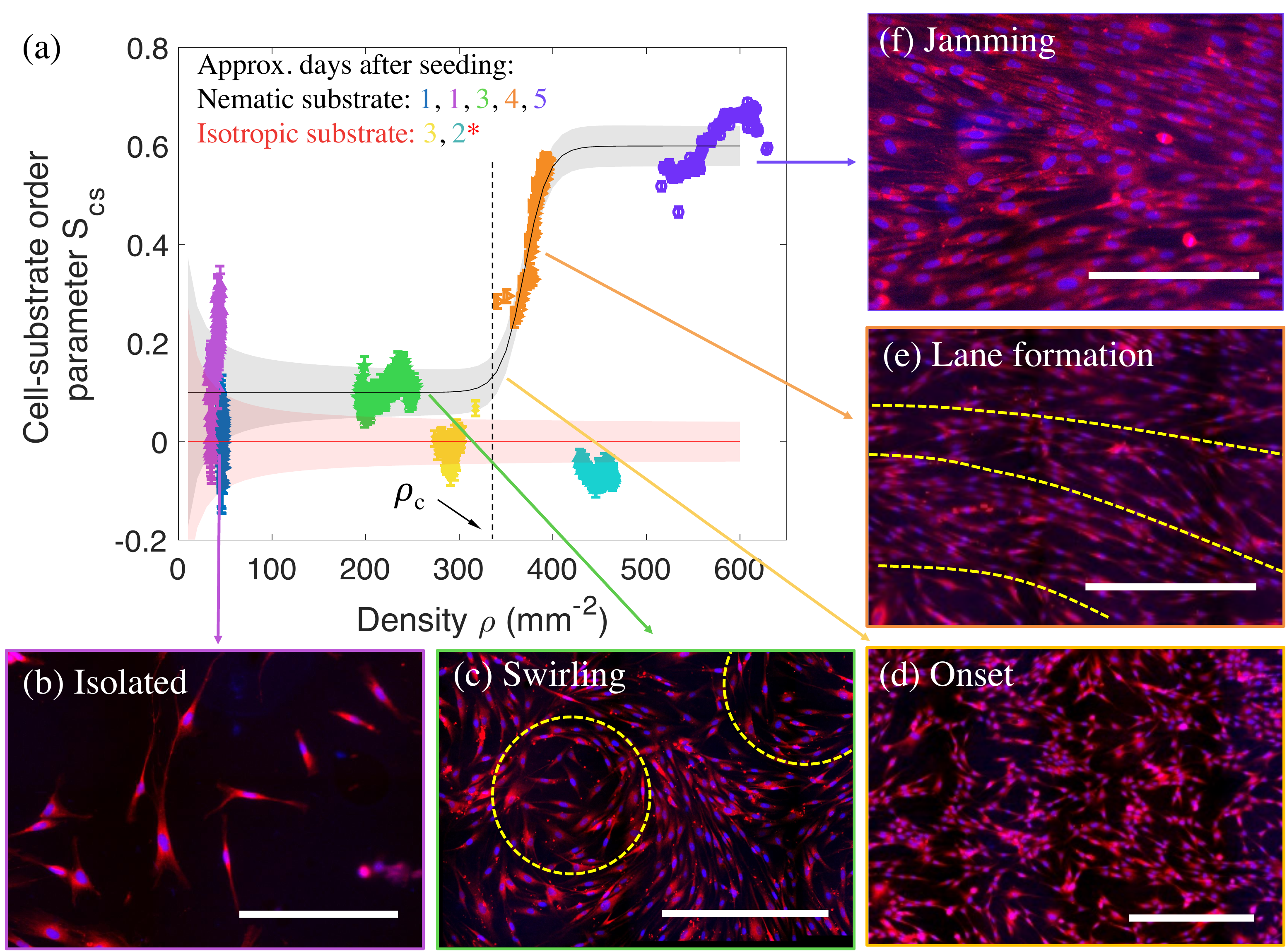}
    \caption{(a) Evolution of the cell-substrate  order parameter $S_{cs}$ with cell density $\rho$. Each color corresponds to a different density obtained  at different times after seeding.  Solid lines serve as  guides to the eye. Shaded regions denote the typical uncertainty in estimating $S_{cs}$ (Fig. \ref{fig:box_size}). Errors in $\rho$ (estimated to be $\sim$ 5-10\%) are not shown. A red asterisk in the legend denotes an experiment with a different initial seeding density. (b) At the lowest densities, cells are isolated and their orientations are random. (c) At low and moderate densities, cells display spontaneous swirling motion. (d) At intermediate to high densities, lanes of cells of coherent velocities form. (e) These lanes widen, and eventually, (f) structural arrest takes place due to jamming. (b-f) are taken on nematic substrates. 
    The scale bars are 250 $\mu$m in (b)(f) and 500 $\mu$m in (c)-(e). }
    \label{fig:order_vs_density}
\end{figure}

(i) At low cell density ($\rho \lesssim$  100 mm$^{-2}$) cells  rarely interact with each other (Fig. \ref{fig:order_vs_density}b) and the monolayer is orientationally disordered. On both nematic and isotropic  substrates individual cells move back and forth in place along their long axis, traveling less than their body length before changing direction (SI Fig. \ref{fig:vacf}, 
Movie S1). With increasing cell density, cells begin to co-localize and one observes the emergence of aligned cell clusters. The spatial fluctuations in cell number $\Delta N$ for given system size $\langle N \rangle$ are greater than those in an equilibrium system  (Fig. \ref{fig:GNF}),  consistent with so-called giant number fluctuations \cite{chate2006simple}, indicating that cell activity contributes to local clustering.  These clusters  nucleate   local orientational order and eventually grow into aligned domains.

(ii) At intermediate densities ($\rho \approx$  200-300 mm$^{-2}$), the monolayer exhibits locally aligned domains, but no global order (Fig. \ref{fig:order_vs_density}a green data, Movie S2). Cell-cell interactions cause individual cells to co-align, which has previously been attributed to steric effects \cite{lober2015collisions} and contact-induced inhibition of locomotion \cite{desai2013contact}. In addition,  we find that on both nematic and isotropic substrates cells tend to follow each other and move along trails. This is demonstrated by the observation of strong  correlations between the velocity of a cell and that of  adjacent leading and trailing cells (SI Fig. \ref{fig:velocity_corr}).  Similar behavior has been observed in surface-dwelling bacteria {\it M. xanthus} \cite{thutupalli2015directional}, and has been  attributed in part to the slime trail they secrete as they glide across a surface.  Fibroblasts are also known to deposit extracellular matrix (ECM) proteins \cite{malakpour2021identification}, and in the current experiments, the presence of fibronectin was confirmed by anti-fibronectin staining (Fig. \ref{fig:fibronectin}), providing evidence for fibronectin as a chemoattractant. This tendency of cells to follow in each other's trail leads to the formation of thin (1-2 cells wide) files of moving cells that do not cross each other and spontaneously organize in swirling patterns (Fig. \ref{fig:order_vs_density}c). 

On nematic substrates, the cell trails tend to  stretch in the direction of substrate nematic order, and begin to resemble ordered lanes of aligned cells that travel in both directions along the long axis of the lane (Movie S3), as observed in simulations of dry active nematics ~\cite{ngo2014large}. In contrast, on isotropic substrates, these swirling patterns persist over time until the cells eventually jam.
To test that the emergence of aligned lanes can be attributed to cell-substrate interaction, we added a small amount (1 $\mu$M) of focal adhesion kinase inhibitor (FAKi)  to the growth media right before imaging  cells on a nematic substrate (Fig. \ref{fig:FAKi_vel}). We found that as a result of this perturbation cell trajectories became more tortuous,  the cell density ceased to grow (Fig. \ref{fig:FAKi_vel}b) and $S_{cs}$ only increased slowly (Fig. \ref{fig:FAKi_vel}c), similar to what was obtained on an isotropic substrates. These observations confirm that molecular-level cell-substrate interactions indeed play a significant role in controlling cell alignment. 

(iii) At even higher densities ($\rho \approx$  300-500 mm$^{-2}$), the collective cell behavior is strongly affected by the properties of the substrate. On isotropic substrates we observe nematically ordered  domains of about 10-15 cells, but there is no global nematic order at the tissue scale. This is evident from measurements of the cell-cell order parameter $S_{cc}$ shown in Fig. \ref{fig:cell_cell_OP}, as well as from the correlation function of cell orientation shown in Fig.~\ref{fig:orientation_corr}a. The cell-substrate order parameter $S_{cs}$ remains zero at all densities as the system has no preferred orientation (Fig.~\ref{fig:cell_cell_OP}a). We also stress that on isotropic substrates cell proliferation becomes arrested and the cell density saturates to a lower value than on nematic substrates.     On nematic  substrates, in contrast, first single-cell files merge to form multi-cell lanes, which align with the direction of substrate nematic order. Lanes of aligned cells moving in both directions coexist with disordered regions, suggesting a phase separation scenario as observed in simulations of dry active nematics~\cite{chate2020dry}. Upon further increase of the density,
both $S_{cc}$ and $S_{cs}$ increase rapidly, tracking each other (see Fig.~ \ref{fig:cell_cell_OP} of SI), and the system transitions to a state of global nematic order on the scale of the entire substrate.  This is also evident from the spatial correlation functions of cellular orientation shown in Fig.~\ref{fig:orientation_corr}a.
The onset of order is estimated to occur roughly at $\rho_c$ = 320 mm$^{-1}$ (Fig. \ref{fig:order_vs_density}d).

\subsection*{A nematic susbstrate drives long-range order of cell orientation}

To shed light on how both cell-cell and cell-substrate interactions work in concert to promote global alignment on the tissue scale, we examined the spatial cell-cell pair orientational correlation function, defined as 
\begin{equation}
    C_{\theta\theta}(r) = \langle2\cos^2 \theta_{ij}(r)-1\rangle.
    \label{eq:orientational_corr}
\end{equation}
where $r$ is the center-to-center distance between cell $i$ and $j$ ($i \neq j$). In practice, we average all pairs with separations within [$r$, $r+dr$] ($dr$ = 10 $\mu$m). The  behavior of $C_{\theta\theta}(r)$ at large $r$ depends qualitatively on whether or not the substrate is aligned. 

\begin{figure}
    \centering
    \includegraphics[width=0.8\textwidth]{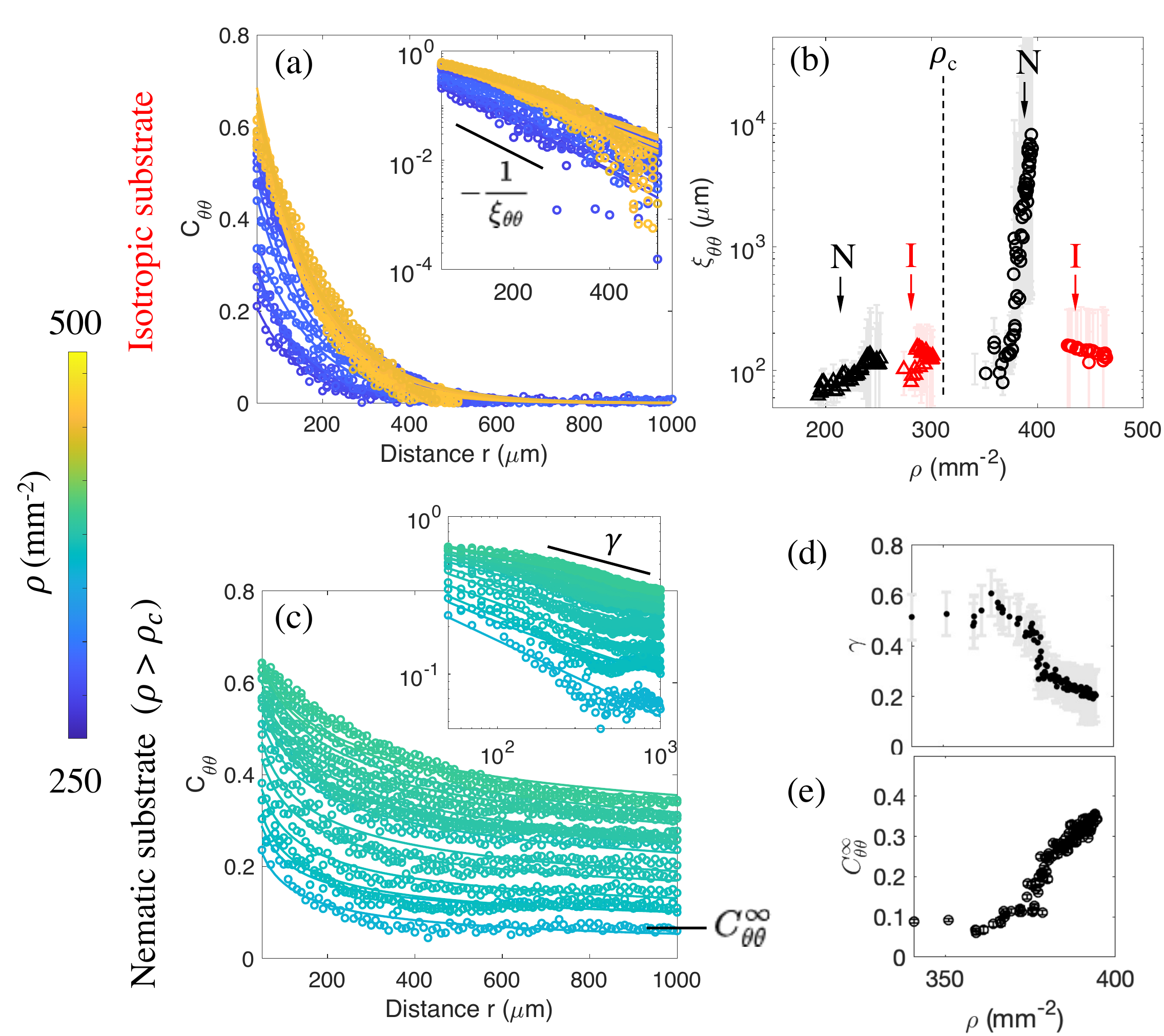}
    \caption{Analysis of spatial correlations of cell-cell orientation. The cell-cell orientational correlation function is plotted as a function of the separation between  cell pairs for (a) isotropic and (c) nematic substrates.
    The insets show the corresponding semilog and log-log plots in (a) and (c), respectively. The correlation length $\xi_{\theta\theta}$ extracted either by fitting Eq. \ref{eq:orientation_corr} or Eq. \ref{eq:modified_orientation_corr} is shown in (b) where red and black symbols denote data obtained using isotropic or nematic substrates, respectively. The yellow vertical line indicates the cell density at the onset of ordering, $\rho_c$. Long-range order is observed for cells on nematic substrates for $\rho > \rho_c$, as demonstrated by the decreasing power law exponent $\gamma$ in (d) and increasing  plateau value $C^\infty_{\theta\theta}$ in (e).  }
    \label{fig:orientation_corr}
\end{figure}

On isotropic substrates $C_{\theta\theta}$ decays exponentially to 0 at all densities (Fig. \ref{fig:orientation_corr}a), demonstrating that in this case cells only have  short-range orientational order \cite{garcia2015physics,yu1998exponential}. A fit to an exponential
\begin{equation}
    C_{\theta\theta}(r) = A_e\exp\left(-\frac{r}{\xi_{\theta\theta}}\right),
     \label{eq:orientation_corr}
\end{equation}
allows us to extract the   orientational correlation length $\xi_{\theta\theta}$ that represents the average size of an  aligned domain 
(Fig.~\ref{fig:orientation_corr}b).

On nematic substrates, by contrast, $C_{\theta\theta}$ plateaus to a non-zero value at large $r$ for  $\rho>\rho_c$, suggesting long-range order on the scale of the system (Fig. \ref{fig:orientation_corr}c). Note that in the absence of the external symmetry-breaking field provided by the aligned substrate, active nematics in two-dimensions are expected to show at most  quasi-long-range order~ \cite{shankar2018low}. Here the substrate acts like an external field and drives the system to a state of global order. Fig. \ref{fig:orientation_corr}e shows the value of $C_{\theta\theta}$ at the largest scale probed in the experiments averaged over the last 15 values, revealing  a clear increase with density.

At intermediate length scales the slow decay of $C_{\theta\theta}$ is well described by a power-law fit, as expected for quasi-ordered  phases in 2D \cite{kosterlitz1973ordering,berezinskii1971destruction}: 
\begin{equation}
    C_{\theta\theta}(r) = A_p r^{-\gamma},
    \label{eq:modified_orientation_corr}
\end{equation}
where $A_p$ is a normalizing constant to ensure $C_{\theta\theta}(r)\leq 1$ for nearest cell pairs, typically located at $r_{min} \approx$ 20 $\mu$m from each other, and $\gamma$ is the power-law exponent. We found that $\gamma$ decreased with increasing $\rho$, reaching a value of 0.1 at the highest density, consistent with the development of long-range order. In this case
we define the correlation length as the distance where $  C_{\theta\theta}(r) $ decays to $1/e$ of its value at $r=r_{min}$,
\begin{equation}
     C_{\theta\theta}(r=\xi_{\theta\theta})=\frac{C_{\theta\theta }(r_{min})}{e}.
    \label{eq:convert_equiv}
\end{equation}
The detailed fitting procedure and parameter estimates are shown in SI Fig. \ref{fig:fit_orientation_corr}.

The fitting reveals that
$C_{\theta\theta}(r_{min})$ is of order 1 and  relatively insensitive to $\rho$ (Fig. \ref{fig:fit_orientation_corr}). This suggests that cells are well-aligned with their nearest neighbors, and their interaction is driving short-range order.
This is supported by the fact that for  $\rho < \rho_c$,  $\xi_{\theta\theta}$ is approximately 100-200 $\mu$m on both substrates  (Fig. \ref{fig:orientation_corr}b). For $\rho > \rho_c$,  $\xi_{\theta\theta}$ increases dramatically when nematic substrates are used, while it remains constant on isotropic substrates. 
Together, these results quantitatively support the conclusion that cells form small aligned domains due to steric repulsion and organize into a state of long-range nematic order when cultured on nematic substrates.

Further evidence that nematic order is driven by steric effects as opposed to correlations in cell velocity, as observed in epithelia, is provided by the computation of spatial velocity correlation functions (see SI Section \ref{sec:pair_velocity}), which are found to decay to zero within $1-2$ cell widths at all densities on both isotropic and nematic substrates (SI Fig. \ref{fig:velocity_corr}a). The distribution of the angle between velocities of of cell pairs is, however, distinctly different. The distribution is isotropic on isotropic substrates, supporting the randomness of cell motion, and bimodal on nematic substrates, with peaks at pair angles of $0$ and $\pi$, confirming the observation of cells' antiparallel motion along aligned lanes (SI Fig. \ref{fig:velocity_corr}c,d). We also observe a strong anisotropy of cell speed on nematic substrates, with cells typically moving almost twice as fast in the direction of nematic alignment than in the transverse direction (SI Fig. \ref{fig:velocity_corr}f,g).

\begin{figure}
    \centering
    \includegraphics[width=0.7\textwidth]{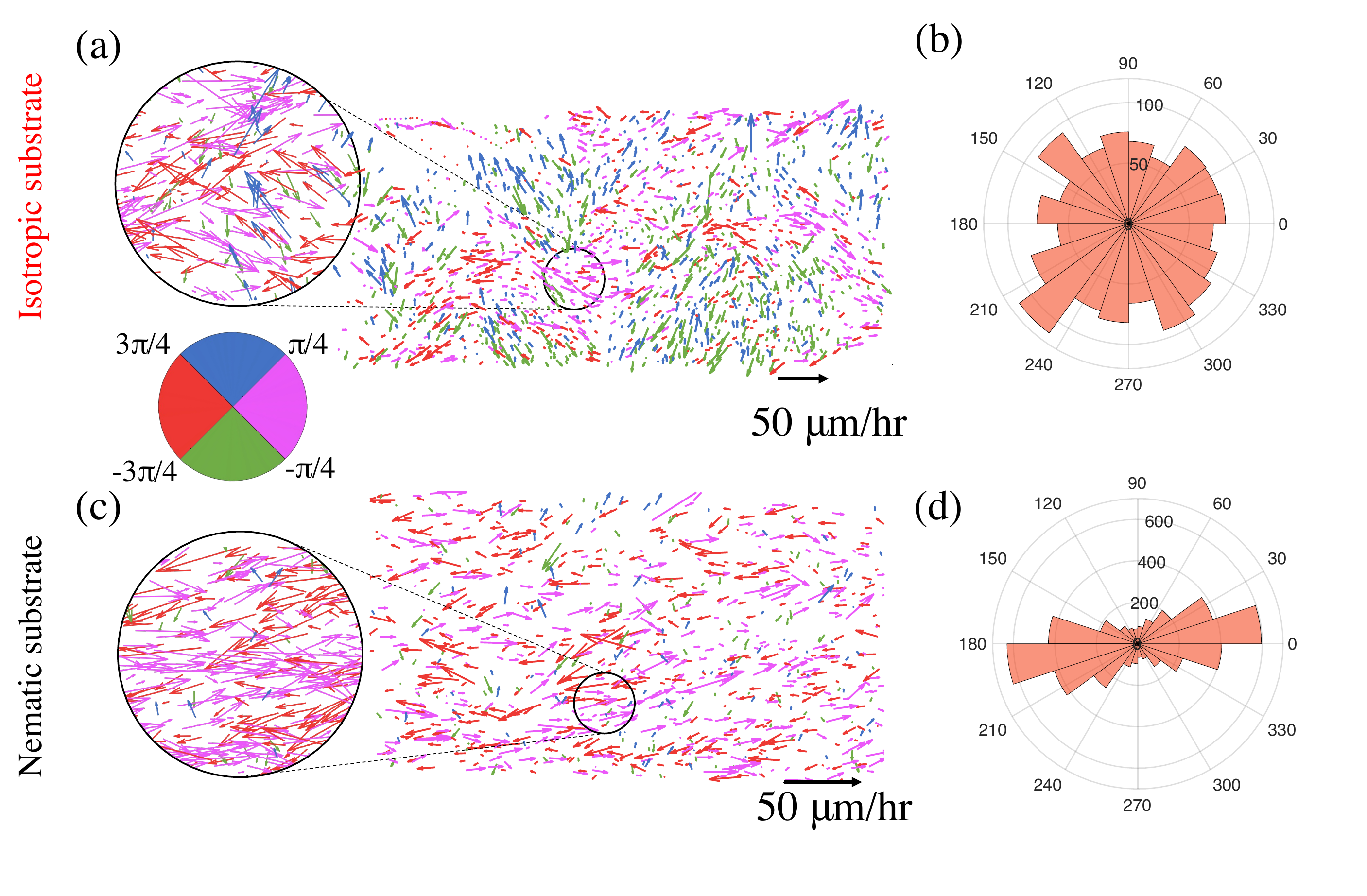}
    \caption{Asymmetric dynamics of cell motion. Spatial maps of the cell velocity field, color-coded by the instantaneous direction of each velocity vector, for cells moving on (a) isotropic and (c) nematic substrates. For clarity, the velocities are plotted at 20\% density in the full FOV, whereas the data of 100\% density is plotted in the zoomed-in figure. The angular distribution of velocity with respect to $\hat{\mathbf e}_x$ is shown  for (b) isotropic and (d) nematic substrates.}
    \label{fig:spatial_vel}
\end{figure}

To further examine these effects, we mapped velocities for a single snapshot close to the jamming densities, as shown in Figure \ref{fig:spatial_vel}, where each cell displacement is represented by a vector (arrow) whose length represents its magnitude, and that points in the direction of instantaneous cell motility.  The arraows are color-coded by binning along the four cardinal directions.  The polar histograms in Fig. \ref{fig:spatial_vel}b,d further demonstate that while the velocity direction is nearly evenly distributed over all angles for cells moving on isotropic substrates (Fig. \ref{fig:spatial_vel}b), cells moving on substrates with nematic order preferentially orient along the substrate alignment direction, here given by ${\pm \hat e_x}$ (Fig. \ref{fig:spatial_vel}d).  Taken together, these observations demonstrate that both steric effects and substrate alignment, which steers cells to move back and forth along bidirectional ``highways'' (SI Fig. \ref{fig:spatial_vel}), are responsible for global nematic order.

Finally, the key role of steric repulsion in driving alignment is also demonstrated by the fact that the experimentally obtained transition density,  $\rho_{c} \approx $ 320 mm$^{-2}$, corresponding to a packing fraction of $\phi \approx 0.5-0.6$, agrees favorably with the transition density for a system of ellipsoid of aspect ratio 6 (similar to the aspect ratio of hdF  cells)~\cite{cuesta1990monte,tan20212d}.

\subsection*{Oriented cell proliferation promotes cell-substrate alignment}

\begin{figure}
    \centering
    \includegraphics[width=0.75\textwidth]{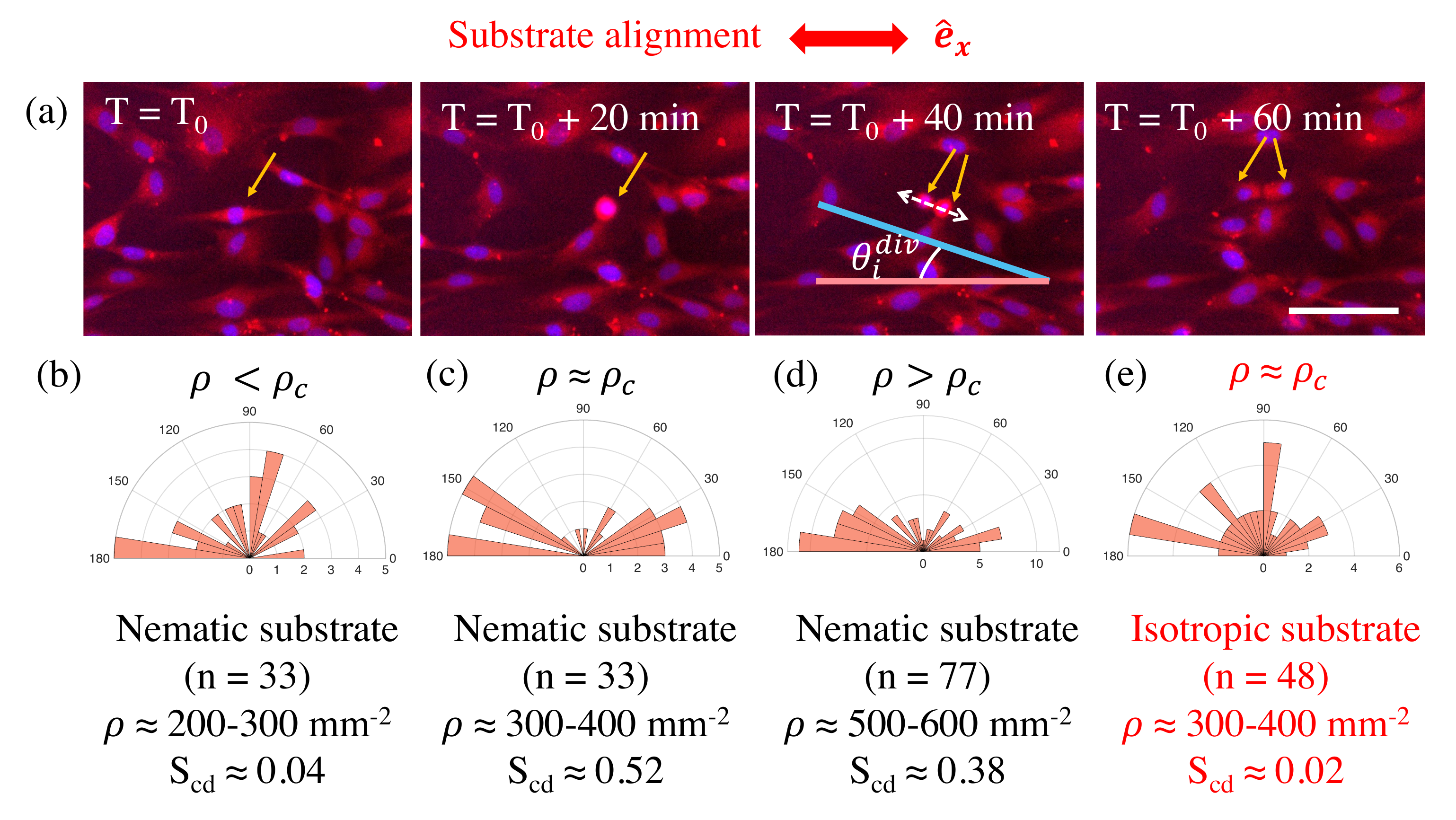}
    \caption{Angular distribution of the cell division axis. (a) The time sequence of images captures a cell-division event. Here $\theta_i^{div}$ denotes the cell-cell orientation at the first time point after the division occurs. (b-e) The polar histograms denote the distribution of $\theta_i^{div}$ for nematic substrate in (b-d) and isotropic substrate in (e). $n$ is the number of observations.}
    \label{fig:cell_division}
\end{figure}

Cell proliferation appears to play an important role in enhancing cell alignment. To quantify this, we tracked  cell division events  at different $\rho$ (Fig. \ref{fig:cell_division}) and calculated the cell division axis order parameter $S_{cd} = \langle 2\cos^2\theta_i^{div} -1 \rangle$, 
where $\theta_i^{div}$  denotes the orientation of the line joining the nuclei of a dividing mother-daughter cell pair at the first time point after division relative to $\hat{\mathbf e}_x$ (Fig. \ref{fig:cell_division}a). For $\rho < \rho_c$, $S_{cd} \approx 0$  on both isotropic (Fig. \ref{fig:cell_division}e) and nematic (Fig. \ref{fig:cell_division}a) substrates.   

On nematic substrates at $\rho \approx \rho_c$,  $S_{cd} \approx 0.52$ (Fig. \ref{fig:cell_division}c).   These results indicate that on nematic substrates cells divide more readily along the substrate orientation direction.  Similarly oriented cells tend to co-localize into larger aligned domains, widening  the lanes and  straightening  them in the alignment direction (Fig. \ref{fig:order_vs_density}e). Cell proliferation may contribute to increased cell-substrate alignment $S_{cs}$ by (1) adding more aligned cells, but also indirectly by (2) making more room so cells are not contact-inhibited and potentially by (3) setting up a extensile dipolar stresses that serve to align nearby cells \cite{doostmohammadi2015celebrating}. Anisotropic proliferation patterns appear to result from weak cell sensing of the substrate's molecular anisotropy. In fact, cell alignment and proliferation appear to reinforce each other until jamming (Fig. \ref{fig:order_vs_density}f) occurs, as also seen in growing bacteria colonies~\cite{volfson2008biomechanical,dell2018growing}. 

To directly test the hypothesis that cell alignment is driven by cell proliferation, we seeded cells at different initial densities $\rho_s$ and fixed them after one week for analysis. For $\rho_s \approx$ 50, 100, and 200 mm$^{-2}$, similar values of $S_{cs} \approx $ 0.68, 0.55, and 0.6 were obtained upon jamming. In contrast, seeding cells at densities larger than $\rho_c$ only leads to partial alignment. For instance, when seeding cells at $\rho_s\approx$ 250, 300, 350 mm$^{-2}$, the monolayers achieved $S_{cs} \approx $ 0.25, 0.45, and 0.13. The degree of final order is likely correlated with the configuration of the cells when first attached. The dependence of cell density and motility on initial seeding cell  density has been noted previously in other experiments, e.g., scratch assays  \cite{jin2016reproducibility}.  We hypothesize that for sufficiently high seeding densities the orientational order can become frustrated because cells cannot efficiently rearrange; as cells become contact inhibited, they cannot divide preferentially along the alignment direction (Fig. \ref{fig:cell_division}) to promote further alignment. 

These phenomena on nematic substrates stand in stark contrast to those of cells on isotropic substrates, where $S_{cs} \approx 0$ for all densities. In this case, cells fail to create long-range order even at high densities, and instead become contact inhibited near $\rho_c$, which leads to the arrest of both proliferation and motion. Experimentally, we measured the cell density at jamming to be $\approx$ 300-450 mm$^{-2}$ for isotropic substrates, as compared to $\approx$  600 mm$^{-2}$ for cells on nematic substrates. Persistent cell motion contributes to local alignment as it effectively enhances the aspect ratio of the cells, and contributes to more efficient packing in the aligned case. While it is possible to seed cells onto an isotropic substrate at higher densities, at $\rho_s \approx$ 300-400 mm$^{-2}$, we found that in the 42 hours of imaging, cell growth remained limited (cyan data in Fig. \ref{fig:order_vs_density}a), compared to the case of a nematic substrate during the same amount of time (purple data in Fig. \ref{fig:order_vs_density}a), despite the cells remaining attached and motile.

\subsection*{Defect movement hints at the nature of intercellular forces}

\begin{figure*}
    \centering
    \includegraphics[width=\textwidth]{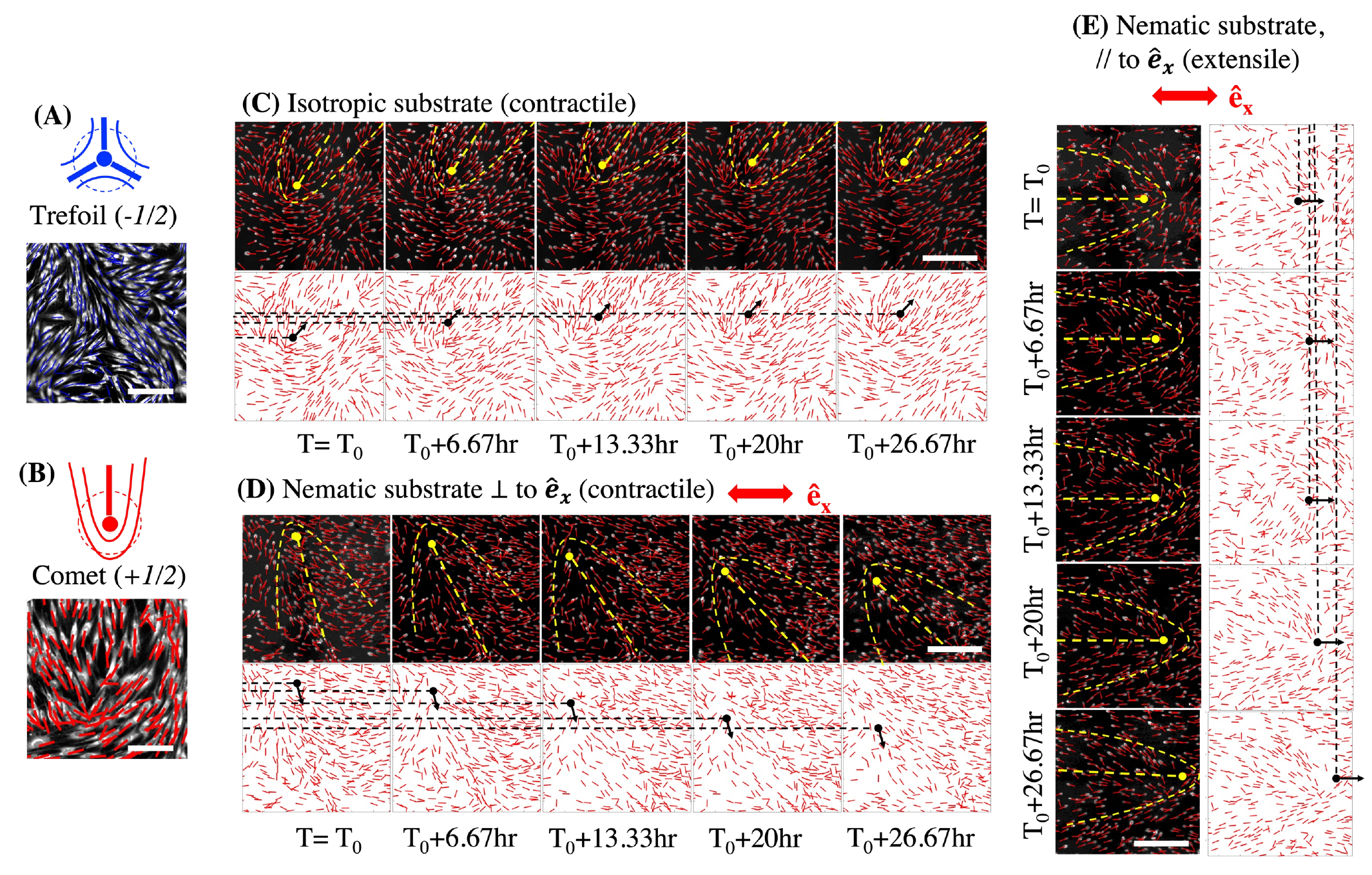}
    \caption{(a-b) shows the (a) trefoil and (b) comet defects. Local director fields are overlaid on the cytoplasm channel. Tracking $+\frac{1}{2}$ defects on (c) isotropic, and on (d-e) nematic substrate, when the comet is (d) perpendicular or (e) parallel to the alignment direction. Red double-sided arrows denote the director field of the substrate. Left and top panels: Orientation overlaid on the nucleus channel, which provides information about the location and orientation of the cell nuclei. Yellow dotted lines denote the approximate outline of the comet and the defect. Right and bottom panels: precise local orientation was determined from the orientation of cell nuclei. The black dots and corresponding arrows denote location and direction of movement of the defects. Dashed lines are added to guide the detection of defect movement. 
    The scale bars are 200 $\mu$m in (a-b), and 100 $\mu$m in (c-e).}
    \label{fig:moving_defect}
\end{figure*}


On both isotropic and nematic substrates, we observed topological defects in the local or global nematic order, as seen in \cite{bade2018edges,kemkemer2000elastic}. On nematic substrates, opposite-sign defects tend to annihilate due to cell motility and tissue remodeling, whereas defects become trapped on isotropic substrates as the system jams, resulting in an arrested glassy state~\cite{angelini2011glass}. In models of 2D nematics,  topological defects can have topological charge $-\frac{1}{2}$ (trefoil, Fig. \ref{fig:moving_defect}a) or $+\frac{1}{2}$ (comet, Fig. \ref{fig:moving_defect}b). In active monolayers, the former is generally stationary, while the latter is motile \cite{giomi2013defect,balasubramaniam2021investigating}. The direction of migration of $+\frac{1}{2}$ defects are controlled by the nature of dipolar active forces in the monolayer: contractile active stresses drive the comet-like defect to move towards its tail, while extensile stresses drive it to move in the direction of the head of the comet~\cite{giomi2013defect,pismen2013dynamics}.

To examine the possible presence and behavior of such defects, we examined high-magnification images of cells at $\rho>\rho_c$ and manually identified and tracked $+\frac{1}{2}$ defects within the hdF monolayer, where the orientation of nematic order is determined by the orientation of the cell nucleus.  We found that on isotropic substrates (Fig. \ref{fig:moving_defect}c and Movie S4) $+1/2$ defects migrated towards their tail, consistent with contractile forces and as previously reported for NIH3T3 fibroblasts \cite{duclos2017topological}. In contrast, on nematic substrates, the direction of defect motion  depended on the defects' orientation with respect to the axis of substrate alignment $\hat{\mathbf e}_x$. Defects oriented perpendicular to $\hat{\mathbf e}_x$ retracted in time, and tended to rotate to align their axis with $\hat{\mathbf e}_x$, suggesting that their motion is again controlled by local contractile forces (Fig. \ref{fig:moving_defect}d, and Movie S5). On the other hand, defects  oriented parallel to $\hat{\mathbf e}_x$ moved in the direction of the head of the comet, indicating that their motion was driven by the extensile stresses arising form oriented cell division (Fig. \ref{fig:moving_defect}e, and Movie S3). Positive defects moving along the boundaries of aligned lanes tended to annihilate with their negative counterparts, enhancing order.
These observations further support the notion that cell division plays an important role in driving nematic order.
Overall, defects are rare on nematic substrates, and we were only able to verify a small number of instances of $+\frac{1}{2}$ defect motion.

\section*{Discussion}

We have shown that the growth and structural organization of hdF cells is controlled by the interplay of spontaneous alignment due to steric cell-cell interactions and the molecular structure of the substrate.  By developing the ability to collect and analyze the dynamics of thousands of cells at high space and time resolution, we established the dynamics of how the cell layer evolves as a function of time as cells proliferate and organize in a nematic state.  On isotropic substrates, cells align with increasing cell density due to crowding, forming locally oriented domains.  The system remains, however, disordered at the global scale, with exponential decay of orientational correlation functions. In contrast, molecular alignment of the LCE substrate provides a preferred alignment direction that allows the monolayer to achieve global nematic order at the tissue scale.
By computing the correlations between cell number, position, velocity and orientation, we identified a sequence of regimes of increasingly rich structure: first the formation of locally oriented domains due to steric effects,  then the emergence of bidirectional lanes where cells travel unobstructed, and finally global orientational order of the entire millimeter-scale tissue. 

Our approach is distinct from other approaches to direct cell monolayer ordering that use surface undulations to drive cellular alignment ~\cite{babakhanova2020cell,turiv2020topology,shin2022recyclable}, where such undulations provide strong guidance. 
Although the molecular mechanisms by which cells sense the substrate molecular alignment remain elusive, there is evidence that cell division occurs preferentially along the direction of substrate alignment and that cell proliferation is enhanced on ordered substrates. It is then tempting to speculate that directed division provides extensile active stresses and an effective anisotropic noise that anneal the monolayers, allowing defects that are essentially jammed on disordered substrates to move and annihilate, resulting in global order. 
 Although individual cells do not respond strongly to the weak alignment cues imposed by the substrate (Fig. \ref{fig:order_vs_density}b), once multicellular lanes form (Fig. \ref{fig:order_vs_density}e), their response to the alignment is enhanced. 
 This enhancement may be influenced by anisotropic tracks of extracellular matrix proteins deposited by other cells, possibly reinforced over time \cite{li2017mechanism}, not unlike what is observed in myxobacteria  \cite{thutupalli2015directional}. More work will, however, be needed to establish the origin of the substrate-induced alignment.

By tracking the movement of spontaneously generated defects, we determined that active stresses in the monolayer appear to be contractile for defects   oriented perpendicular to the substrate alignment direction (Fig. \ref{fig:moving_defect}d), but extensile  when the defect is oriented parallel to the direction of substrate alignment (Fig. \ref{fig:moving_defect}e).
Such extensile stresses are likely to arise from oriented cell division along the direction of substrate alignment. In epithelial layers,  a uniaxial stretch of the substrate can drive preferentially oriented cell-division along the stretch direction, though E-cadherin seems  to be critical in transducing the stress and regulate cell-division orientation \cite{hart2017cadherin}. Our study suggests that hdFs not only sense the local substrate orientation and align their long axes to the substrate alignment direction, but  also actively remodel the cell monolayer at the tissue level  \cite{saraswathibhatla2020tractions}. This sensitivity and behavior may play an important role in the organization of newly deposited ECM proteins in connective tissue, or in the development of multicellular aggregates in biological contexts. Taken together, these observations suggest new avenues for the development of guiding surfaces to direct wound healing or tissue regeneration in biomedical engineering applications.

The development of methods to reliably control and program collective cellular alignment and motility is essential for use in tissue engineering and medical applications. Our work shows that a nematically ordered, but topologically flat substrate can serve as an external field to drive global order. It additionally uncovers the evolution of spatial and temporal correlations during the transition to the ordered state, revealing a remarkable connection to well-studied models of dry active nematics. Finally, it demonstrates new techniques for rapid collection and analysis of large data sets of tissue-scale structure and dynamics, which inform mechanistic insights and lay the experimental groundwork for future theoretical and data science studies aimed at elucidating the molecular mechanisms of cell-substrate interaction that drive collective alignment.
  


\section*{Method}

\subsection*{A high-throughput imaging and single cell tracking framework to study cell dynamics} 

To investigate the emergence of nematic order in a model cell monolayer, we cultured hdF cells on LCE substrates with either nematic or isotropic molecular structures (Fig. \ref{fig:substrate}), both having identical chemical composition. Substrates were fabricated following previous protocols \cite{martella2019liquid} using a mixture of monoacrylates (RM23), diacrylates (RM82), and trace amounts of photo-initiator (Fig. \ref{fig:substrate}a). The ratio of RM82 : RM23 = 1:1 (mol/mol), was tuned to produce substrates with appropriate mechanical properties and alignment capability. The polymer mixture was heated above its melting temperature and introduced into a thin chamber consisting of two glass slides pre-coated with a polyvinyl alcohol solution, separated by thin spacers. In cases where uniform nematic alignment of the LCE film was desired, the slides were rubbed unidirectionally with a velvet cloth. Solid films (Fig. \ref{fig:substrate}b) of nematic or isotropic order were obtained by crosslinking the polymer at a temperature either below or above the isotropic-nematic transition point (for full details, see SI, Section \ref{sec:materials}, \ref{sec:lce}). Alignment was confirmed by examining the substrates between crossed polarizers (Fig. \ref{fig:dsc}b-c).

The degree of orientational order within the LCE substrate was determined by 2D wide angle X-ray diffraction (WAXD, Fig. \ref{fig:WAXS}). The substrate mechanical properties were characterized by dynamic mechanical analysis (DMA, Fig. \ref{fig:dma}), which showed an approximately three-fold increase in elastic modulus parallel to the alignment direction relative to the perpendicular direction, whereas the elastic modulus of the isotropic substrate was uniform and had an intermediate value between the two. Importantly, the film processing method assured that the substrates were flat and smooth, which in turn ensured that at low cell density, there were no significant topographical cues to drive the alignment of individual cells. This was experimentally confirmed by the small ($< 2$ nm) root-mean squared roughness measured using atomic force microscopy, and this result was insensitive to the direction of probing (parallel or perpendicular to the LCE alignment direction) (Fig. \ref{fig:confocal_alignment}a). We further confirmed the flatness by mapping the surface topography using scanning electron microscopy (Fig. \ref{fig:confocal_alignment}b). 

Cells were cultured using standard protocols (SI Section \ref{sec:cell_handling}) and seeded onto clean substrates without further chemical modification.  Substrates were aligned to the imaging chamber such that the nematic alignment direction was along $\mathbf{\hat e_x}$. Before imaging, the cytoplasm and nuclei were stained with fluorescent markers following standard protocols (SI Section \ref{sec:imaging}).  The adherent, motile cells have a highly elongated body shape with aspect ratios of $\sim$4-6, and nuclei that are also anisotropic, with an aspect ratio that varies from 1.5 to 2.5. nucleus orientation is highly correlated with cell body orientation (Fig. \ref{fig:substrate}e), having a correlation coefficient of 0.81 (Fig. \ref{fig:nucleus_body_alignment}). Fluorescence signals from the nuclei are well-separated and easier to track than the cytoplasmic ones, particularly at high cell density, and are therefore used to indicate the local orientation of the cell. 

A high-throughput imaging workflow was developed to enable quantitative analysis of cellular dynamics over large length and time scales. First, using a precision motorized stage and automated image collection routines, we acquired scans of tens of fields-of-view (FOVs, Fig. \ref{fig:substrate}c), which were stitched together in postprocessing to obtain a composite image such as Fig. \ref{fig:substrate}d at every time point, each taking less than 2 minutes to scan. Then, nucleus position and orientation, which were used to indicate cell position and orientation, were determined for each image frame using a custom analysis scheme that enabled thresholding on a per image basis to account for bleaching and spatial variations (Fig. \ref{fig:verify_tracking}), and by fitting an ellipse to the detected nucleus region. We represent the nucleus orientation of cell $i$ by the angle $\theta_i$ with respect to $\hat{\mathbf e}_x$, which for the nematic LCE substrates represents the direction  of nematic alignment (Fig. \ref{fig:substrate}e). The positions of the nuclei at subsequent times were then linked to form trajectories (Fig. \ref{fig:cell_traj}). We then determined the cell velocity by computing the cell displacement and dividing it by the  time interval.  Each experiment records the trajectories of a few thousand cells, including cell position, orientation, major and minor axes, velocity, and ID number for each cell at over 100 time points, which can be postprocessed and analyzed in minutes. This method allows for  precise and robust quantification of  cell orientation, density and motility over long times and large fields of view.

\bibliography{cell}

\bibliographystyle{Science}

\section*{Acknowledgments}

This work was supported by the Otis Williams Postdoctoral Fellowship from the Division of Mathematical, Life, \& Physical Sciences in the College of Letter \& Science, with partial support from the BioPACIFIC Materials Innovation Platform of the National Science Foundation under Award No. DMR-1933487 (NSF BioPACIFIC MIP), by Hellman Family Faculty Fellowship from University of California, by the MRSEC Program of the NSF under Award No. DMR-1720256 (IRG-3) and by the U.S. Army Research Office under Cooperative Agreement W911NF-19-2-0026 for the Institute for Collaborative Biotechnologies. Views and conclusions are those of the authors and should not be interpreted as representing official policies, either expressed or implied, of the U.S. Government. MCM acknowledges partial support from the National Science foundation under Award No. DMR-2041459. MG acknowledges partial support from the National Science foundation under Award No. 2053423. XF acknowledges the partial support from UCSB academic senate faculty research grant program. The authors thank Jen Smith for training on cell tissue culture preparation and acknowledge the use of the Biological Nanostructures Laboratory and Microfluidics Laboratory within the California NanoSystems Institute, supported by the University of California, Santa Barbara and the University of California, Office of the President. Authors also acknowledge use of the UC Santa Barbara MRL Shared Experimental Facilities that are supported by the MRSEC Program of the NSF under Award No. DMR-1720256; a member of the NSF-funded Materials Research Facilities Network (www.mrfn.org).


\section*{Supplementary materials}

Position, orientation and velocity extraction procedures

Variable interpretation

Material characterization

Cell handling and imaging

Figs. S1 to S16

Movies S1 to S5

References \textit{(1-8)}


\end{document}